\documentclass{article}
\usepackage{spconf,amsmath, graphicx}
\usepackage{comment}
\usepackage{cite}
\usepackage{color}
\usepackage{enumerate}
\usepackage{array}
\usepackage{breqn}
\usepackage{amssymb}

\usepackage{multirow}
\usepackage{arydshln}
\title{Interpretable Filter Learning Using Soft Self-attention \\ For Raw Waveform Speech Recognition}
\name{Purvi Agrawal and Sriram Ganapathy
 \thanks{This work was partly funded by grants from
  the Department of Atomic Energy (DAE) project (DAEO0205), the Ministry of Human Resource and Development (MHRD), Government of India.}
}
\address{Learning and Extraction of Acoustic Patterns (LEAP) lab,\\ Electrical Engineering, Indian Institute of Science, Bangalore, India.}
\ninept
\begin{document}
\ninept
\maketitle
\begin{abstract}
Speech recognition from raw waveform involves learning the spectral decomposition of the signal in the first layer of the neural acoustic model using a convolution layer. 
In this work, we propose a raw waveform convolutional filter learning approach using soft self-attention. The acoustic filter bank in the proposed model is implemented using a parametric cosine-modulated Gaussian filter bank  whose parameters are learned. A network-in-network architecture provides self-attention to generate attention weights over the sub-band filters. The attention weighted log filter bank energies are fed to the acoustic model for the task of speech recognition. 
Experiments are conducted on Aurora-4 (additive noise with channel artifact), and CHiME-3 (additive noise with reverberation) databases. In these experiments, the attention based filter learning approach provides considerable improvements in ASR performance over the baseline mel filter-bank features and other robust front-ends (average relative improvement of $7$\% in word error rate over baseline features on Aurora-4 dataset, and $5$\% on CHiME-3 database). Using the self-attention weights, we also present an analysis on the interpretability of the filters for the ASR task. 
\end{abstract}

\section{Introduction}
\label{sec:intro}
Even with several advancements in automatic speech recognition (ASR) systems using deep learning \cite{hinton2012deep} and sequence modeling \cite{hochreiter1997long}, there is significant performance degradation in noisy and reverberant environments. 
For most of the speech recognition systems, the first processing step is the extraction of features like mel filter-bank or gamma-tone filter-bank features \cite{mfccdavis, gammatone1987efficient}. This feature extraction step approximates the early part of human hearing. Recently, with the advent of neural networks, feature learning from data has been actively pursued from raw waveform \cite{doss2013, sainath2013, tuske2014acoustic}.

In a supervised data-driven approach, the underlying model can automatically discover features needed for the objective at hand from the raw signals, e.g. detection or classification. Several works like \cite{sainath2013, hoshen2015speech, sainath2015cldnn,ravanelli2018interpretable} have specifically incorporated the learning of acoustic mel-like filters using convolution operations in the first layer of network. Many of these approaches also use mel filter initialization for the first filter-bank layer and the final learned filters have a close similarity to the mel-filters. While the approaches have yielded insights into the data driven filters, the interpretability is  limited. There has been some early attempts to explore interpretability of filters recently \cite{ravanelli2018interpretable}. 

In sequence-to-sequence  modeling tasks like machine translation ~\cite{luong2015effective} and speech recognition~\cite{bahdanau2016end}, the use of network-in-network (NIN) architecture to derive attention weights has provided a significant boost in interpretability of such models~\cite{vaswani2017attention}. For example, the analysis of attention in machine translation can tell if the translation is accurate at a word level in addition to being accurate at a sentence level. In these bidirectional recurrent neural network architectures, the attention network is  provided with a feedback from the output prediction. The modification to self-attention which requires no feedback from the output~\cite{lin2017structured} 
allows the extension of attention framework to all types of  neural architectures~\cite{yu2018qanet}. For example, in tasks like language recognition,  self-attention reveals features that are more relevant to the task~\cite{padi2019end}. 
The self-attention was also introduced for speech recognition in  \cite{dong2018selfAttention_transformer, povey2018time}. The self-attention networks have the ability to establish direct dependencies between any layer in the network with the targets~\cite{dong2018selfAttention_transformer}.

In this paper, we hypothesize that representation learning can be efficiently performed with self-attention based filter-bank weighting approach. This work proposes a soft self-attention weighting approach applied on the output of the first layer of a deep model. The first layer performs acoustic filter-bank learning from the raw waveform using a convolutional layer. The acoustic filters are parametric cosine-modulated Gaussian filters~\cite{agrawal2019unsupervised} whose parameters are learned within the acoustic model. The convolution is carried out in time domain, and the output of the layer is pooled and log transformed to obtain time-frequency representation. The output is also fed to the NIN module to obtain self-attention weights for the filter-bank outputs. The weighted filter-bank representation is fed to the neural network architecture for the task of speech recognition. All the model parameters including the acoustic filter learning layer and the self-attention weights are learned in a supervised learning paradigm. The filter weights are initialized using the unsupervised learning framework~\cite{agrawal2019unsupervised}.

The ASR experiments are conducted on Aurora-4 (additive noise with channel artifact) and CHiME-3 (additive noise with reverberation) databases. The experiments show that the learned representations from the proposed  framework of filter learning with self-attention provides considerable improvements in ASR results over the baseline mel filter-bank features and other robust front-ends. We  analyze the attention weights provided by the self-attention layer in the trained deep model. 
We also investigate the performance of the proposed framework in a semi-supervised setting where availability of labeled data is limited.
The rest of the paper is organized as follows. Sec. \ref{sec:rep_learning} describes the proposed representation learning approach. Sec.~\ref{sec:experiments} describes the ASR experiments with the various front-ends followed by the results and analysis. We conclude the work with summary in Sec. \ref{sec:summary}.
\begin{center}
    \begin{figure}[t]
        \centering
        \includegraphics[trim={0.12in 2.6in 00in 0.72in}, clip, scale=0.415]{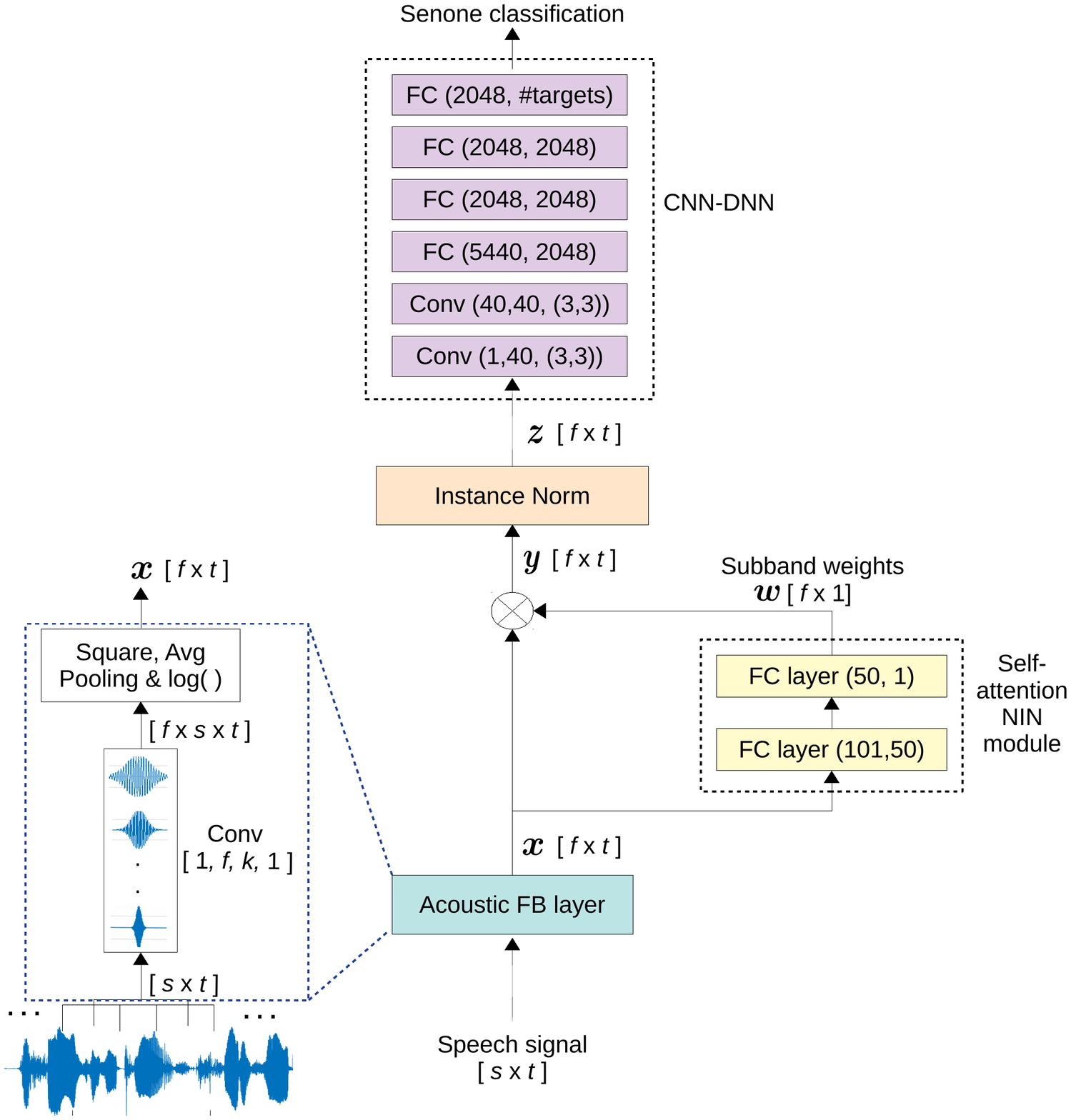}
        \vspace{-0.5cm}
        \caption{Block diagram of the proposed acoustic filter learning with soft self-attention for ASR acoustic modeling.}
        \label{fig:block_diag}
        \vspace{-0.2cm}
    \end{figure}
\end{center}

\vspace{-1.3cm}
\section{Attention Based Acoustic Filter Learning}{\label{sec:rep_learning}}
The block schematic of the proposed soft self-attention based acoustic filter learning model is shown in Fig.~\ref{fig:block_diag}.

\subsection{Acoustic Filter-bank learning} 
 The first layer of the proposed ASR model performs acoustic filtering  learnt from the raw waveforms using a convolutional layer.  The input to the neural network are raw samples windowed into $s$ samples per frame with a contextual window of $t$ frames. This matrix of size $s \times t$ raw audio samples are processed with a 1-D convolution  using $f$ kernels ($f$ also denotes the number of sub-bands in filter-bank decomposition) each of size $k$.  The kernels are modeled as cosine-modulated Gaussian function \cite{agrawal2019unsupervised},
\vspace{-0.05cm}
\begin{equation}
    {{w}}_i (n) = \cos{2\pi\mu_i n} \times \exp{(-{n^2}\mu_i^2/{2})}
    \vspace{-0.05cm}
\end{equation}
where ${{w}}_i (n)$ is the $i$-th kernel ($i=1,..,f$)  at time $n$, $\mu_i$ is the center frequency of the $i$th filter (in frequency domain), and variance of the Gaussian is tied to the mean as $\sigma_i = 1/\mu_i$. The number of filter taps is denoted as $k$.  The parametric approach to FB learning generates filters with a smooth frequency response. We initialize the means $\mu _i$ through unsupervised pre-training using convolutional varational autoencoder (CVAE)  \cite{agrawal2019unsupervised}. 

The convolution with the cosine-modulated Gaussian filters generates $f$ feature maps. These outputs are squared, average pooled within each frame and log transformed. This generates $\boldsymbol{x}$ as $f$ dimensional features for each of the $t$ contextual frames, as shown in Fig. \ref{fig:block_diag}. 

\begin{figure}[t]
    \centering
    \includegraphics[trim={0 0 0 0}, clip, scale=0.3]{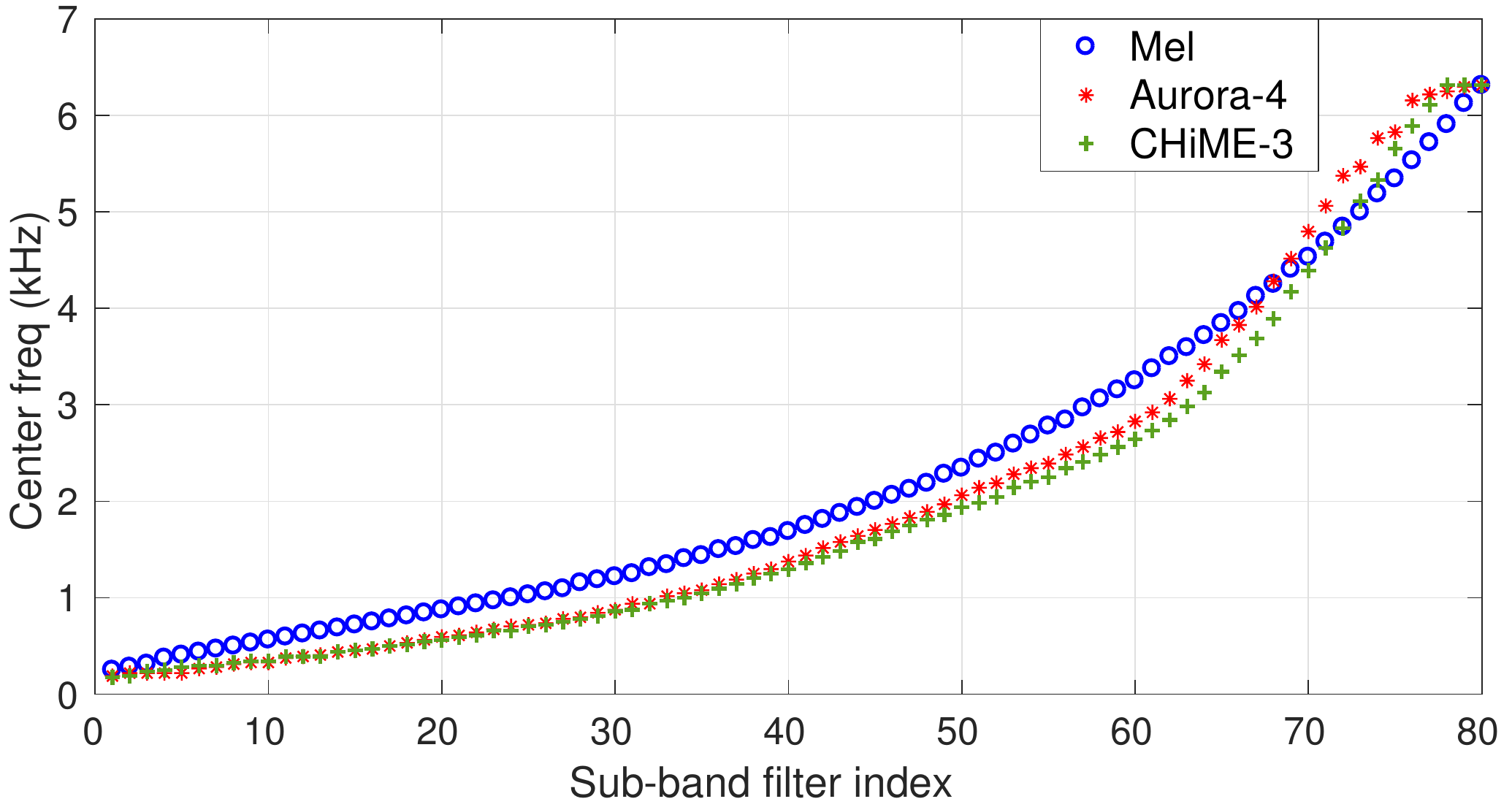}
    \vspace{-0.2cm}
    \caption{Comparison of center frequency of acoustic filterbank learnt with proposed approach for Aurora-4 and CHiME-3 datasets, with center frequencies of mel filterbank.}
    \label{fig:center_freq_acoustic}
     \vspace{-0.3cm}
\end{figure}
\subsection{Soft Self-attention module} {\label{subsec:self_attention}}
The attention paradigm is implemented using a network-in-network (NIN) module fed with the $f \times t$ output of the acoustic filter-bank layer from the previous step. The two layer DNN network with a softmax output generates weights $\boldsymbol{w}$ as $f$ dimensional vector with weights corresponding to each filter. 

The attention weights over the $f$ dimensional features tend to be small values for many sub-bands and results in hard suppression of sub-band features that are critical for phoneme separation. In order to overcome this issue, we propose a soft attention scheme applied on the attention weighted filter-bank outputs $\boldsymbol{y}$. This is inspired by instance norm principle~\cite{rumelhart1986learning, ulyanov2016instance}. Let $y_{j,i}$ denote the attention weighted filter-bank output for frame $j$ ($j=1,..,t$) of sub-band $i$ ($i=1,..,f$). The soft attention output $z_{j,i}$ is given as,
\vspace{-0.15cm}
\begin{equation}\label{eq:soft}
    z_{j,i} = \frac {y_{j,i} - m_{i}} {\sqrt{\sigma ^2 _i + c}} 
    \vspace{-0.15cm}
\end{equation}
where $m_i$ is the sample mean of $y_{j,i}$ over $j$ and $\sigma _i$ is the sample std. dev. of $y_{j,i}$ over $j$. The constant $c$ acts as a relevance factor. When the attention weight for sub-band $i$ is high, the std. dev. $\sigma _i$ is also high compared to $c$ and thus the soft attention output $z_{j,i}$ has a unit variance over $j$. When the attention weight for sub-band $i$ is low, the value of $\sigma _i$ is also relatively less compared to $c$ and this makes the variance of $z _{j,i}$ lower than $1$. Thus, Eq.~\ref{eq:soft} modulates the attention mechanism and provides a soft version of the attention weights to be propagated for the acoustic modeling in ASR. 

Following the acoustic filter-bank layer and the self-attention NIN module, the acoustic model consists of series of CNN and DNN layers. The configuration details are given in Fig.~\ref{fig:block_diag}. 
In our experiments, we use $t=101$ whose center frame is the triphone target for the acoustic model. We also use $f=80$ sub-bands with $k=129$. This value of $k$ corresponds to $8$ ms in time for a $16$ kHz sampled signal. The value of $s$ is $400$ corresponding to $25$ ms window length and the frames are generated at $10$ms shifts.  Thus, the input to the acoustic filter bank layer is about $1$ sec. of audio. In our experiments, we also find  that after the instance norm layer, the number of frames $t$ can be pruned to the center $21$ frames alone for the acoustic model training without loss in performance. This has significant computational benefits and we prune the output of the soft-attention to keep only the $21$ frames around the center frame ($200$ ms of context). 

Fig. \ref{fig:center_freq_acoustic} shows the center frequency ($\mu_i$ values sorted in ascending order) of the acoustic filters obtained using multi-condition Aurora-4 and CHiME-3 datasets (details of the datasets are given in Sec. \ref{sec:experiments}) and this is compared with the center frequency of the mel filterbank. As can be observed, the proposed filterbank has more number of filters in lower frequencies compared to the mel filterbank. 

\begin{figure}[t]
    \centering
    \includegraphics[trim={0in, 0 0in 0in}, clip, scale=0.33]{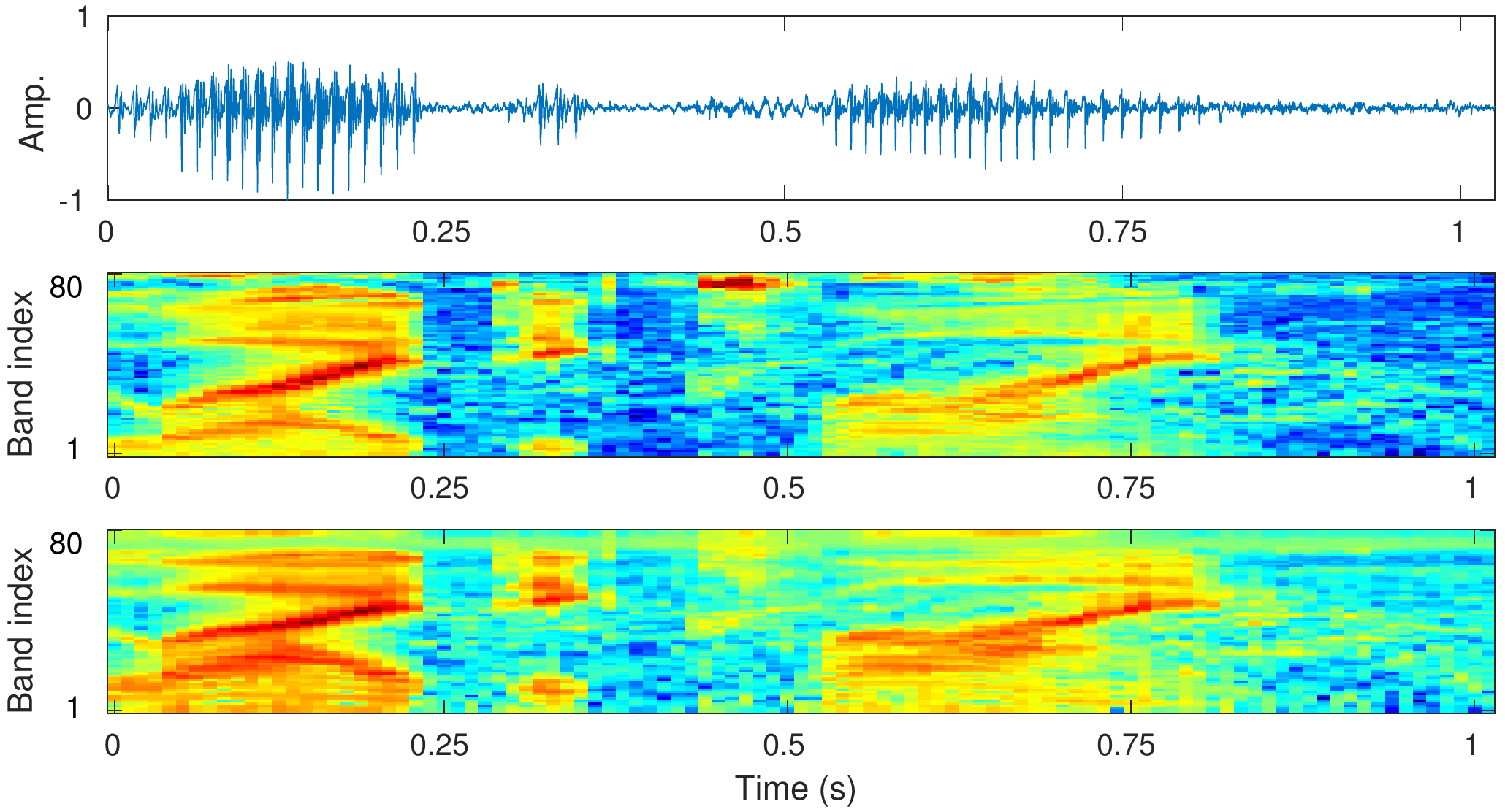}
    \vspace{-0.2cm}
    \caption{(a) Speech signal from Aurora-4 dataset with airport noise, (b) mel spectrogram representation without attention (c) acoustic filterbank representation with soft self-attention ($\boldsymbol{z}$ in Fig. \ref{fig:block_diag}). }
    \label{fig:sig_spec}
    \vspace{-0.4cm}
\end{figure}

The soft self-attention weighted time-frequency representation $\boldsymbol{z}$ obtained from the proposed approach is shown in Fig. \ref{fig:sig_spec}(c) for an utterance with airport noise from Aurora-4 dataset (the waveform is plotted in Fig. \ref{fig:sig_spec}(a)). The corresponding mel spectrogram with instance normalization (without attention) is also plotted in Fig. \ref{fig:sig_spec}(b). It can be observed that in the proposed approach, the formants appear to be shifted upwards because of the increased number of filters in the lower frequency region. Also, the attention weighting helps to reduce the effect of noise in higher sub-bands.
\vspace{-0.2cm}
\section{Experiments and results}{\label{sec:experiments}}
The speech recognition system is trained using PyTorch \cite{paszke2017pytorch} while the Kaldi toolkit~\cite{povey2011kaldi} is used for decoding and language modeling. The ASR is built on two datasets, Aurora-4 and CHiME-3 respectively. The models are discriminatively trained using the training data with cross entropy loss and Adam optimizer \cite{kingma2014adam}. A hidden Markov model - Gaussian mixture model (HMM-GMM) system trained using  with MFCC features is used to generate the triphone alignments for training the CNN-DNN based model. The ASR results are reported with a language model re-scoring of the lattices, where the lattices generated with tri-gram language model are rescored using recurrent neural network language model (RNN-LM) \cite{mikolov2011rnnlm} for the final ASR decoding. The best language model weight is obtained from development set. For each dataset, we compare the ASR performance of the proposed approach of filter-bank learning with soft attention (Raw-Att) with traditional mel filter-bank energy (MFB) features, power normalized filter-bank energy (PFB) features \cite{kim2012pncc}, RASTA features (RAS) \cite{hermanskyb}, and mean Hilbert envelope (MHE) features \cite{mhec2015}. All the features are processed with CMVN on a 1 sec running window. The architecture shown in Fig. \ref{fig:block_diag} (except for the acoustic filterbank layer and the attention module) is used for all the baseline features.

\begin{center}   
     \begin{table}[t]
     \vspace{-0.4cm}
            \centering
            \begin{center}
            \caption{Word error rate (\%) in Aurora-4 database for multi-condition training with various feature extraction schemes.}
            \label{tab:multiData_aurora4_rnnlm}
            \resizebox{8.6cm}{3.9cm}{
            \begin{tabular}{|l|c|c|c|c|c|c|c|}
            \hline
            Cond & MFB & PFB  & RAS & {MHE} & Raw & MFB-Att & Raw-Att \\ \hline
            \multicolumn{8}{|c|}{A. Clean with same Mic} \\ \hline
            Clean & 3.4 & 3.4 & 4.2 & 3.1 & 3.2 & 3.5 & \textbf{2.9} \\ \hline
            \multicolumn{8}{|c|}{B: Noisy with same Mic} \\ \hline
            Airport & 6.0 & 6.3 & 6.8 & 6.3 & 5.2 & 5.9 & \textbf{5.1}  \\
            Babble & 6.1 & 6.4 & 7.4  & 6.5 & 5.6 & 6.3 & \textbf{5.2}  \\
            Car & {3.5} & 4.0 & 4.2 & 3.8  & 3.6 & 3.6 & \textbf{3.4}  \\
            Rest. & 8.4 & 8.2 & 9.5 & 8.0 & 7.3 & 8.2 & \textbf{6.8}  \\
            Street & 6.8 & 7.1 & 8.3 & 7.0 & 6.9 & 7.0 & \textbf{6.1}  \\
            Train & 7.4 & 7.4 & 8.5 & 7.4 & 7.1 & 7.4 & \textbf{6.4}  \\\hdashline
            Avg. & 6.4 & 6.6 & 7.4 & 6.5 & 5.9 & 6.4 & \textbf{5.5} \\
             \hline
            \multicolumn{8}{|c|}{C: Clean with diff. Mic}\\ \hline
		    Clean & 6.1 & 6.2 & 7.6 & 6.4 & 6.1 & \textbf{6.0} & {6.9} \\ \hline
            \multicolumn{8}{|c|}{D: Noisy with diff. Mic}  \\ \hline
            Airport & 14.9 & 16.9 & 16.1 & 16.8 & 15.7 & 15.5 & \textbf{14.6}  \\
            Babble & 15.5 & 16.7 & 18.1 & 16.8 & 16.4  & 15.6 & \textbf{14.8}   \\
            Car & {7.7} & 9.8 & 9.0 & 8.2 & 7.9 & \textbf{7.4} & {8.3}  \\
            Rest. & 17.0 & 19.5 & 19.9 & 18.3 & 17.1 & 17.3 & \textbf{16.2}  \\
            Street & 16.0 & 17.9 & 17.6 & 17.6 & 16.7 & 16.6 & \textbf{15.8}  \\
            Train\ & 16.3 & 17.3 & 18.2 & 17.1 & 16.7 & 16.5 & \textbf{15.2}  \\ \hdashline
            Avg. & 14.6 & 16.3 & 16.5 & 15.8 & 15.1 & 14.8 & \textbf{14.1}  \\
             \hline
            \multicolumn{8}{|c|}{Avg. of all conditions}  \\ \hline
            Avg. & 9.7 & 10.5 & 11.1 & 10.2 & 9.7 & 9.8 & \textbf{9.1} \\ \hline
            \end{tabular}
            }
          \end{center}
       \end{table}
   \end{center}
\vspace{-0.3in}

\subsection{Aurora-4 ASR}
The WSJ Aurora-4 corpus is used for conducting ASR experiments. This database consists of continuous read speech recordings of 5000 words corpus, recorded under clean and noisy conditions (street, train, car, babble, restaurant, and airport) at $10-20$ dB SNR. The training data has 7138 multi condition recordings (84 speakers) respectively. The validation data has 1206 recordings for multi condition setup. The test data has 330 recordings (8 speakers) for each of the 14 clean and noise conditions. The test data is classified into group A - clean data, B - noisy data, C - clean data with channel distortion, and D - noisy data with channel distortion.

The ASR  performance for the proposed (Raw-Att) features (soft self-attention on acoustic filterbank representation as discussed in Sec. \ref{sec:rep_learning}) is shown in Table \ref{tab:multiData_aurora4_rnnlm} for each of the 14 test conditions. For Aurora-4 dataset, we also compare the ASR performance with the acoustic filterbank representation (Raw) without attention. In addition, we learn and apply the self-attention weights over MFB features (MFB-Att) for ASR.

As seen in the results, most of the noise robust front-ends do not improve over the baseline mel filterbank (MFB) performance. The Raw waveform features perform similar to MFB baseline features on average while performing better than the baseline for Cond. A and B. The MFB-Att features, which constitute the application of the attention layer without filter-bank learning, also doesn't improve over baseline MFB features. The proposed feature extraction scheme combines filter-bank learning with soft attention. These features  provide  considerable improvements in ASR performance over the baseline system with average relative improvements of $7$\% over MFB features. Furthermore, the improvements in ASR performance are consistently seen across all the noisy test conditions except condition C. In particular, the relative improvements in same microphone conditions (A and B) are about $15$\% relative compared to the baseline system. 




\begin{table}[t]
\begin{center}
\caption{Word error rate (\%) in CHiME-3 Challenge database for multi-condition training (real+simulated) with test data from simulated and real noisy environments.}
\label{tab:Chime3Results_rnnlm}
	\begin{tabular}{|l|c|c|c|c|c|c|}
	\hline
		Test Cond & ~MFB~ & ~PFB~ & ~{RAS}~ & ~{MHE}~ & ~Raw-Att~ \\ \hline
		Sim\_dev & 10.1 & 10.6 & 11.7 &  10.1  & \textbf{9.8} \\ 
		Real\_dev & 7.5 & 7.9 & 8.6 & 7.7 & \textbf{{7.3}} \\ \hdashline
		Avg. & 8.8 & 9.2 & 10.1 & 8.9 & \textbf{8.5} \\ \hline \hline

		Sim\_eval & 16.0 & \textbf{15.2} & 18.5 & 15.6 & {15.3} \\ 
		Real\_eval & 14.4 & 15.6 & 16.5 & 14.7 & \textbf{13.6} \\\hdashline
		Avg. & 15.2 & 15.4 & 17.5 & 15.2 & \textbf{14.4} \\ \hline 

	\end{tabular}
\end{center}
\vspace{-0.6cm}
\end{table}


\begin{table}[t]
\begin{center}
\caption{WER (\%) for each noise condition in CHiME-3 dataset with the baseline features and the proposed feature extraction.}
\label{tab:Chime3Results_detailed_rnnlm}
    \resizebox{8.6cm}{1.15cm}{
	\begin{tabular}{|c|c|c|c|c||c|c|c|c|}
	\hline
	 & \multicolumn{4}{c||}{\textbf{Dev Data}} & \multicolumn{4}{c|}{\textbf{Eval Data}} \\\hline
	\multirow{2}{*}{Cond.} & \multicolumn{2}{c|}{Sim} & \multicolumn{2}{c||}{Real} & \multicolumn{2}{c|}{Sim} & \multicolumn{2}{c|}{Real} \\\cline{2-9}
	& MFB & Prop & MFB & Prop & MFB & Prop & MFB & Prop \\\hline
	BUS & \textbf{8.3} & {8.8} & 9.1 & \textbf{{8.8}} & 10.5 & \textbf{10.4} & \textbf{17.4}  & {18.0}\\
	CAF & 13.7 & \textbf{12.4} & \textbf{7.2} & {7.3} & 18.1 & \textbf{16.6} & 14.2 & \textbf{13.1} \\
	PED & 7.8 & \textbf{7.8} & 5.6 & \textbf{5.6} & 16.3 & \textbf{14.8} & 14.3 & \textbf{11.9}\\
	STR & 10.8 & \textbf{10.3} & 8.0 &\textbf{7.3} & \textbf{19.2} & {19.3} & 11.5 & \textbf{11.5} \\ \hline
	\end{tabular}
	}
\end{center}
\vspace{-0.7cm}
\end{table}
\begin{figure*}
    \centering
    \includegraphics[trim={0 0 0in 0}, clip, scale=0.355]{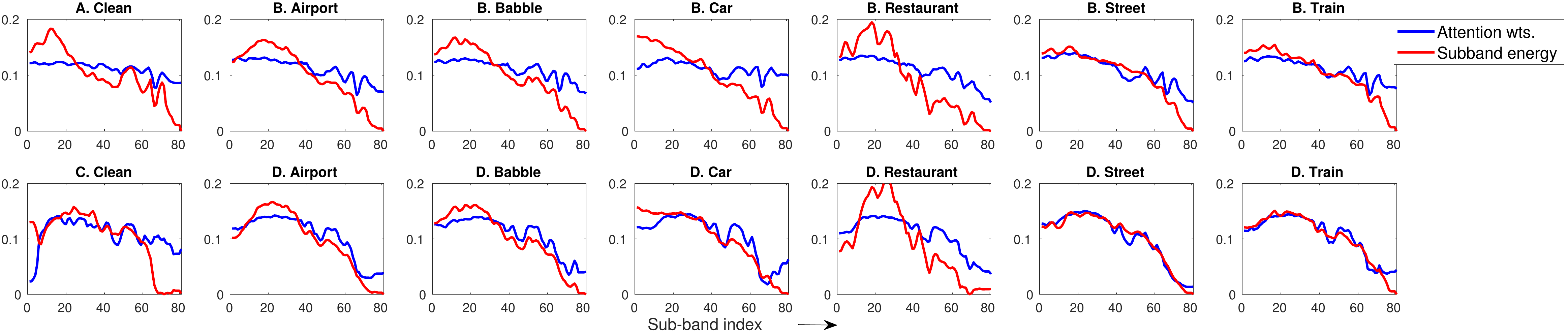}
    \vspace{-0.2cm}
    \caption{The plots show the average  self-attention weights for a random  utterance from 14 different test conditions in blue color and the corresponding average sub-band energy profile in red color. The attention weights follow the signal-to-noise ratio in the input signal and tend to provide lower weights to low signal energy regions. }
    {\label{fig:noisy_feat_avg}}
     \vspace{-0.5cm}
\end{figure*}
\subsection{{CHiME-3 ASR}}
The CHiME-3 corpus for ASR contains multi-microphone tablet device recordings from everyday environments, released as a part of 3rd CHiME challenge \cite{barker2015chime3}. Four varied environments are present, cafe (CAF), street junction (STR), public transport (BUS) and pedestrian area (PED). For each environment, two types of noisy speech data are present, real and simulated. The real data consists of $6$-channel recordings of sentences from the WSJ$0$ corpus spoken in the environments listed above. The simulated data was constructed by artificially mixing clean utterances with environment noises. The training data has $1600$ (real) noisy recordings and $7138$ simulated noisy utterances.
We use the beamformed audio in our ASR training and testing. The development (dev) and evaluation (eval) data consists of $410$ and $330$ utterances respectively. For each set, the sentences are read by four different talkers in the four CHiME-3 environments. This results in $1640$ ($410 \times 4$) and $1320$ ($330 \times 4$) real development and evaluation utterances in total. Identically-sized, simulated dev and eval sets are made by mixing recordings captured in the recording booth with the environmental noise recordings.

The results for the CHiME-3 dataset are reported in Table \ref{tab:Chime3Results_rnnlm}. The proposed  approach of raw waveform filter learning with soft attention provides considerable improvements over the baseline system as well as the other noise robust front-ends considered here. On the average, the proposed approach provides relative improvements of 5\% over MFB features in the eval set. The detailed results on different noises in CHiME-3 are reported in Table \ref{tab:Chime3Results_detailed_rnnlm}. For most of the noise conditions in CHiME-3 in simulated and real environments, the proposed approach provides improvements over the baseline features.

\vspace{-0.3cm}
\subsection{Semi-supervised training}
\begin{center}
    \begin{figure}[t!]
        \centering
        \includegraphics[trim={0 0 0 1.6cm}, clip, scale=0.32]{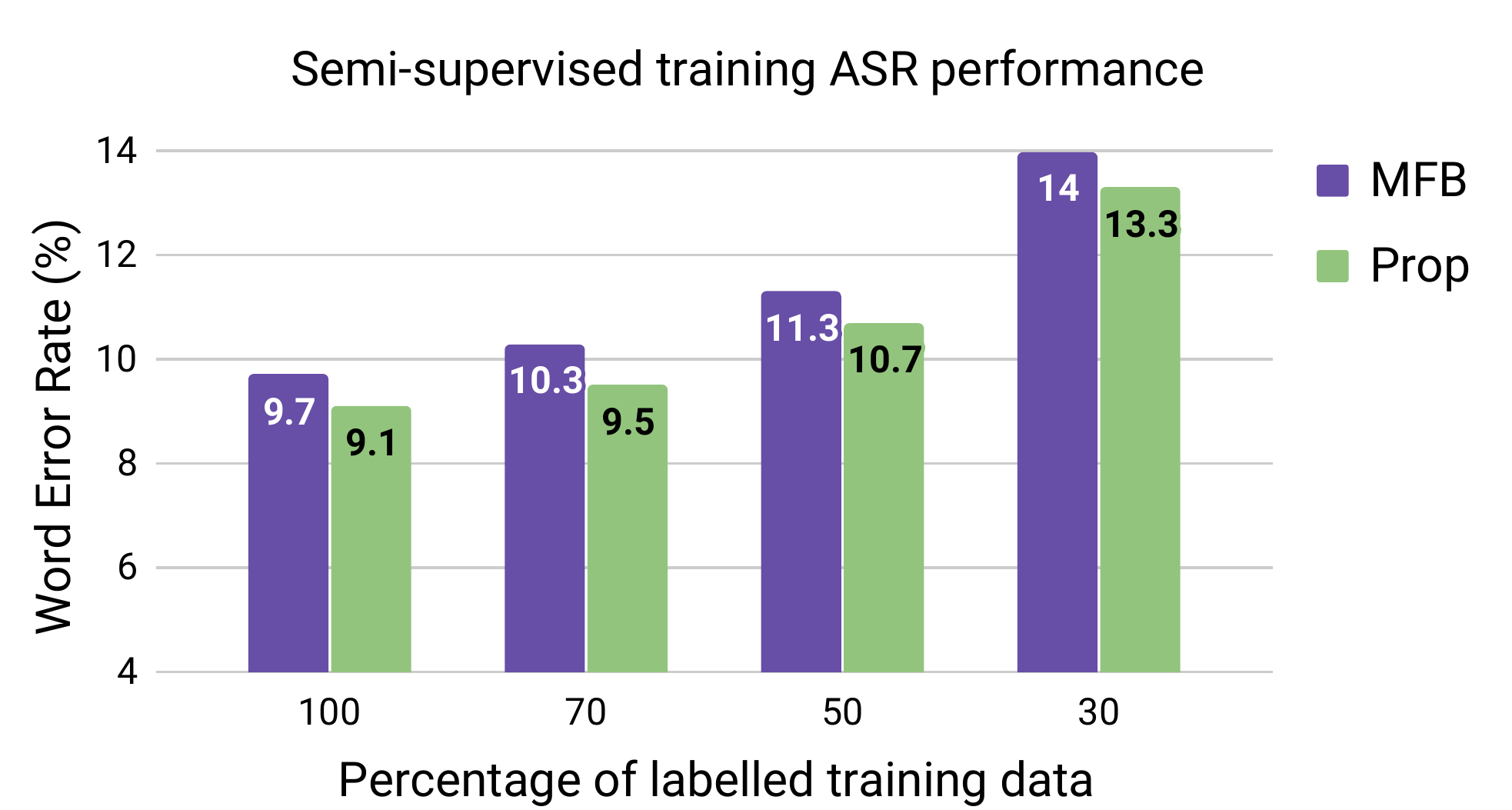}
        \vspace{-0.3cm}
        \caption{ASR performance in WER (\%) for Aurora-4 database (avg. of 14 test conditions) using lesser amount of labeled training data.}
        \label{fig:semisup}
        \vspace{-0.4cm}
    \end{figure}
\end{center}

{\label{sec:sem-sup}}
In this section, we test whether the filter-bank learning with soft attention is robust to the lack of supervised training data.  We consider the case when only a fraction of the available training data is labeled. This is partly motivated by the fact that, while data collection in real noisy environments may be relatively easy, the labeling of noisy data is cumbersome and more expensive than in clean recording conditions. For semi-supervised ASR training, the Aurora-4 training set up is used with $70$, $50$ and $30$\% of the labeled training data. The performance comparison of ASR with semi-supervised training is shown in Fig. \ref{fig:semisup} for MFB and the proposed Raw-Att approach. As seen here, the proposed approach consistently performs better than the baseline MFB features  even when the amount of labeled training data is small. These experiments show that the filter bank learning framework is not data hungry and the filter parameters can be learned effectively with limited supervised data. 

\subsection{Discussion}{\label{sec:discussion}}
We analyze the soft self-attention weights for the Aurora-4 test data under various noise conditions. Fig. \ref{fig:noisy_feat_avg} shows the average attention weights of the sub-bands for an utterance of Aurora-4 dataset from 14 different test noise types in blue color. We also plot the corresponding sub-band energy profile of acoustic filterbank representation (averaged over all frames for an utterance) from all 14 test conditions in red color. Both the attention weights and the sub-band energies are unit length normalized in the plot.

From the plot, it can be observed that the obtained attention weights correlates with the sub-band energy profiles in most of the test conditions. The sub-band energies have more magnitude in the lower sub-band region (sub-bands $1 - 40$) as compared to the higher sub-bands. The attention weights also follow similar trend for most of the test conditions, except for clean test condition, where we observe that the attention weights are almost flat. 
Thus, the attention weights provide information to the ASR to deweight the sub-band regions that are low in energy and vulnerable to noise.

\vspace{-1cm}
\section{Summary}{\label{sec:summary}}
The major contribution of the work are as follows:
\begin{itemize}
    \item Proposed an interpretable filter learning approach using soft self-attention from raw waveform.
    \item  The acoustic filter bank in the first convolutional layer of the proposed model is implemented using a parametric cosine-modulated Gaussian filter bank whose parameters are learned.
    \item A network-in-network architecture provides self-attention to obtain attention weights over the sub-band filters. 
    \item The proposed attention based feature learning for ASR gives considerable improvements in multiple datasets over baseline features. The performance improvements are consistent in semi-supervised ASR training as well.
    \item Analysis of the attention weights shows that it correlates well with the signal energy profile in the sub-bands.
\end{itemize}

\vfill\pagebreak

\bibliographystyle{IEEEbib}
\bibliography{refer,refs}

\end{document}